\documentclass[12pt,a4paper,english]{article}
\usepackage{mathptmx}
\usepackage[T1]{fontenc}
\usepackage[latin9]{inputenc}
\usepackage{amsmath}
\usepackage{graphicx}
\usepackage{setspace}
\usepackage{esint}
\makeatletter

\newcommand{\lyxaddress}[1]{
\par {\raggedright #1
\vspace{1.4em}
\noindent\par}
}
\newenvironment{lyxlist}[1]
{\begin{list}{}
{\settowidth{\labelwidth}{#1}
 \setlength{\leftmargin}{\labelwidth}
 \addtolength{\leftmargin}{\labelsep}
 }}
{\end{list}}

\makeatother

\makeatother

\usepackage{babel}
\begin{document}

\title{A comparative study of aqueous DMSO mixtures by computer simulations
and integral equation theories}

\author{Aurélien Perera$^{1}$ and Bernarda Lovrin\v{c}evi\'{c}$^{2}$ }
\maketitle

\lyxaddress{$^{1}$Laboratoire de Physique Théorique de la Matière Condensée
(UMR CNRS 7600), Université Pierre et Marie Curie, 4 Place Jussieu,
F75252, Paris cedex 05, France.}

$^{2}$Department of Physics, Faculty of Sciences, University of Split,
Nikole Tesle 12, 21000, Split, Croatia.
\begin{abstract}
Several computer simulation studies of aqueous dimethylsulfoxyde with
different force field models, and conducted by different authors,
point out to an anomalous depressing of second and third neighbour
correlations of the water-water radial distribution functions. This
seemingly universal feature can be interpreted as the formation of
linear water clusters. We test here the ability of liquid state integral
equation theories to reproduce this feature. It is found that the
incorporation of the water bridge diagram function is required to
reproduce this feature. These theories are generally unable to properly
reproduce atom-atom distribution functions. However, the near-ideal
Kirkwood-Buff integrals are relatively well reproduced. We compute
the Xray scattering function and compare with available experimental
results, with the particular focus to explain why this data does not
reproduce the cluster pre-peak observed in the water-water structure
factor.
\end{abstract}

\section{Introduction}

The aqueous dimethyl-sulfoxyde(DMSO) mixtures have been the subject
of several studies, mainly because of their peculiar properties at
low temperatures \cite{10Kirshner}. One of the principal interest
concerning this mixture is the nature of the water-DMSO interactions,
in relation to the low temperature behaviour and cryoprotectancy\cite{12cryo}.
Because of this special interest, most of the earlier studies seem
to have focused in the water-DMSO clusters, as even very recent studies
show the same interest\cite{14BorinSkaf,16iceberg,18cluster,20castner,22recent,23recent}.
DMSO has been modeled differently by different authors and there are
numerous subsequent computer simulation studies of these particular
mixtures\cite{30RaoSingh,31VMmodel,32chandlerLuzar2,33chandlerLuzar1,34gunsteren,35hawlicka,36Laaksonen,37allAtom,38nicoDMSO,39luzar,40ladanyi}.
However, despite the various possible water-DMSO force field model
combinations, it would appear that all of them indicate that the water-water
correlations show the same typical feature, namely a depression of
the second and third neighbour correlations. This is illustrated in
Fig.1, where the main panel clearly shows that second and third neighbour
correlations show oscillations around 1, in contrast with that of
aqueous ethanol (dotted line) which is above 1, as usual with aqueous
mixtures. Albeit typical variations due to small differences between
DMSO models, the universality of this feature is quite visible. While
the first peak witness the strong water-water hydrogen bond, which
is seen in any mixtures, the depletion of farther neighbours suggests
that water has to satisfy 2 constraints: being hydrogen bonded to
itself and loose the favoured tetrahedral conformation of bulk water.
One solution to this problem is the formation of water string-like
clusters. Indeed, such clusters are clearly visible in snapshots from
simulations, and at all DMSO mole fractions $x$ in the range $0.1<x<0.9$,
as we have been the first to point out their existence \cite{45ourDMSO1}.
Since various types of experiments show more clearly the existence
of water-DMSO bonds\cite{22recent,23recent}, one could induce that
linear-water clusters must be a consequence of the first more readily
visible property. Indeed, all the force field models of DMSO share
the fact that the oxygen site is charged negatively while all the
other sites, sulfur and the 2 methyl sites, are charged positively.
It is quite easy to visualise that the negatively charged water oxygen
site would bind to these latter DMSO sites, leaving the 2 hydrogen
bonds open to the bulk, hence favouring neighbouring water oxygens
to attach to them. This situation will clearly disfavour tetrahedral
binding of water. This is what computer simulations seem to indicate.
This situation is very different from other aqueous mixtures, as for
example, water-alcohol mixtures, where water is seen to form globular-like
self segregated domains\cite{47waterMH}. In Ref.\cite{45ourDMSO1},
we have argued that it is the large size of the DMSO atom which allows
water oxygen to attach to it. However, another puzzle arises: why
available Xray scattering data for this mixture\cite{50KogaXray}
does not show a pre-peak, which should arise from the existence of
such water clusters. Indeed, the presence of such clusters induces
a clear pre-peak at about $k\approx0.75$\AA$^{-1}$ in the water-water
structure factor that we have reported in Ref.\cite{45ourDMSO1}.
Hence, this pre-peak should, in principle, contribute to the total
scattered intensity. In the present paper, we explain the absence
of scattering pre-peak in narrow connection to the concept of domain
order introduced in Ref.\cite{55myDomOrd}, which leads to a compensation
of the positive pre-peak in like species structure factors by the
negative pre-peak of the cross species structure factors. This is
always the case for many aqueous mixtures, which show the same universal
local domain order. In order to break this ``symmetry'' and induce
a scattering pre-peak, it is necessary that the interface separating
the segregated water and solute domains acquires particular properties,
which would enhance its contribution to the total scattering. This
is the case of micelles, for example, for which water is seen to accumulate
in the coronary area. We conjecture here that this is the reason why
aqueous mixtures of small mono-ols and 1-2-diols do not show scattering
pre-peaks, while these tend to appear in triols, which have ester
groups near the surfactant OH head group, and which attract water
in the corona of the micelle.

Concerning liquid state integral equation theories (IET) approaches,
these have been intensively tested in the early 90's, but have generally
lead to disappointing results for associated liquids\cite{57LueBlanck1,57LueBlanck2}.
To make this point clear, both the molecular IET and the site-site
IET are enable to describe the tetrahedral bonding of water through
the water-water pair correlation function\cite{57LueBlanck2,61a,61b}.
The main point we want to make here is that these theories are even
less successful for mixtures involving associated liquids, precisely
because association induces domain segregation, which is a more complex
property to capture. In particular, we want to show that correcting
the deficiencies of pure liquid description, by introducing bridge
diagram for water, cannot account for the dramatic change in structure
of segregated mixtures. The particular interest to test these theories
for the case of aqueous DMSO, is the fact that this mixtures shows
quasi-ideal Kirkwood-Buff integrals, unlike most other aqueous mixtures\cite{65Matteoli}.
In that, this particular system seems to have less concentration fluctuation,
and therefore should be more amenable to mean-field like approaches
such as IET. In Ref.\cite{45ourDMSO1}, we have argued that this feature
of the aqueous DMSO mixtures was related to water linear aggregates
forming pseudo molecular entities, leading to a weakly interacting
pseudo ideal mixture of DMSO with these aggregates. Since IET are
quiet good for weakly interacting and fully disordered mixtures, it
would be interesting to see how much of this property can be captured
by these theories. Indeed, there are 2 problems here: the first is
low concentration fluctuations, which is favourable for a IET description,
and the second is the description of hydrogen bonded aggregates, which
should be a challenge for IET.

The remainder of this paper is as follows. In the next section, we
briefly remind the models used, simulation protocols and theoretical
details related to the IET and their use in the present case. Details
concerning the concept of domain order and calculation of the scattering
functions are equally gathered in the last subsection. Section 3 details
our results for the structural functions and the KIB. Finally, Section
4 gathers our conclusions and perspective on the present work.

\section{Theoretical details}

\subsection{Models and computer simulations}

We have essentially followed the methodology reported in Ref.\cite{45ourDMSO1}.
Most of the data shown here is from the same simulations presented
in this work. Additional structure factors have been computed in order
to allow for the calculation of the scattering functions. To summarize,
we have used the SPC/E water model\cite{66berendsen} and the VB DMSO
model\cite{66VBmodel}. We have also used in Fig.1 the OPLS DMSO model\cite{67oplsDMSO}.
$N=2048$ particle simulations were used in the isothermal and isobaric
molecular dynamics simulations with the Gromacs package\cite{68gromacs}.
Specific details can be found in Ref.{[}\cite{45ourDMSO1}{]}.

\subsection{Integral equation theory}

We have used the site-site IET formalism, which has the same level
of accuracy as the molecular version, as far as associated liquids
are concerned. Although one would expect a superior accuracy of the
molecular formulation, due to an exact version of the molecular Ornstein-Zernike
(MOZ) equation\cite{69patey,70hansmac}, instead of the approximate
site-site Ornstein-Zernike (SSOZ) version\cite{57LueBlanck1,70hansmac},
it is the constraint imposed by closure relation which dominates the
structural description. Indeed, the description of hydrogen bonded
structures in the associated liquids is particularly constraining,
requiring 3-body and higher order correlations present in the missing
highly connected and irreducible bridge diagrams\cite{70hansmac}.
This is the reason for the similarity in the poor description of both
approaches for water from MOZ\cite{61a} and SSOZ\cite{61b}. 

The 2 equations to solve are the site-site OZ matrix equation together
with a closure equation. The SSOZ equation reads

\begin{equation}
S.M=I\label{SSOZ}
\end{equation}
where

\begin{equation}
S=W+\rho H\label{SF}
\end{equation}
 is the generalized structure factor matrix, with elements defined
as

\begin{equation}
S_{i_{\alpha}j_{\beta}}(k)=W_{i_{\alpha}j_{\beta}}(k)+\rho H_{i_{\alpha}j_{\beta}}(k)\label{Sk}
\end{equation}
where the index $i_{\alpha}$ and $j_{\beta}$ designate atom i of
molecular species $\alpha$ and atom j of molecular species $\beta$
, respectively, and
\begin{equation}
M=W^{-1}-\rho C\label{M}
\end{equation}
The matrix $W$ describes the intramolecular correlation, defined
as rigid atomic bonds within molecular species

\begin{equation}
W_{i_{\alpha}j_{\beta}}(k)=\frac{\sin(kd_{i_{\alpha}j_{\alpha}})}{kd_{i_{\alpha}j_{\alpha}}}\label{W}
\end{equation}
where $d_{i_{\alpha}j_{\alpha}}$ is the distance between atomic sites
$i_{\alpha}$ and $j_{\alpha}$ of molecule of species $\alpha$.
The matrix H is related to the site-site distribution functions $g_{i_{\alpha}j_{\beta}}(r)$,
with $h_{i_{\alpha}j_{\beta}}(r)=g_{i_{\alpha}j_{\beta}}(r),-1$,
and $\tilde{h}_{i_{\alpha}j_{\beta}}(k)=\int d\vec{r}h_{i_{\alpha}j_{\beta}}(r)\exp(i\vec{k}.\vec{r})$
the Fourier transform of $h_{i_{\alpha}j_{\beta}}(r)$, and the matrix
C is related to the Fourier transform of the site-site direct correlation
function $c_{i_{\alpha}j_{\beta}}(r)$ with the relations

\[
H_{i_{\alpha}j_{\beta}}=\sqrt{x_{\alpha}x_{\beta}}\tilde{h}_{i_{\alpha}j_{\beta}}(k)
\]
\[
C_{i_{\alpha}j_{\beta}}=\sqrt{x_{\alpha}x_{\beta}}\tilde{c}_{i_{\alpha}j_{\beta}}(k)
\]
where $x_{\alpha}$ is the mole fraction of species $\alpha$.

For the closure equations, we have chosen both the hypernetted chain
(HNC) equation and the Kovalenko-Hirata (KH) closure\cite{72KH}.
The HNC closure equation reads

\begin{equation}
g_{i_{\alpha}j_{\beta}}(r)=\exp\left(-\beta v_{i_{\alpha}j_{\beta}}(r)+h_{i_{\alpha}j_{\beta}}(r)-c_{i_{\alpha}j_{\beta}}(r)+b_{i_{\alpha}j_{\beta}}(r)\right)\label{HNC}
\end{equation}
where $v_{i_{\alpha}j_{\beta}}(r)$ is the interaction between the
atomic sites $i_{\alpha}$ and $j_{\beta}$ ($\beta=1/k_{B}T$ is
the Boltzmann factor), and the last term $b_{i_{\alpha}j_{\beta}}(r)$
represents the so-called bridge function. Setting $b_{i_{\alpha}j_{\beta}}(r)=0$
leads to the strict HNC closure, while this function can also be tailored
from the site-site correlation functions obtained through the procedure
we have outlined in Ref.\cite{73ourism}. 

The KH equation is a mix between the strict HNC equation above and
the mean spherical approximation, based on the value of the term $\Gamma_{i_{\alpha}j_{\beta}}(r)=-\beta v_{i_{\alpha}j_{\beta}}(r)+h_{i_{\alpha}j_{\beta}}(r)-c_{i_{\alpha}j_{\beta}}(r)$
in the exponential of Eq.\ref{HNC}:
\begin{equation}
g_{i_{\alpha}j_{\beta}}(r)=\begin{cases}
\exp\left(\Gamma_{i_{\alpha}j_{\beta}}(r)\right) & \mbox{when }\Gamma_{i_{\alpha}j_{\beta}}(r)<0\\
1+\Gamma_{i_{\alpha}j_{\beta}}(r) & \mbox{when }\Gamma_{i_{\alpha}j_{\beta}}(r)>0
\end{cases}
\end{equation}

This ad-hoc procedure of damping the raise of the correlations in
the exponential through the term $\Gamma_{i_{\alpha}j_{\beta}}(r)$
allows to control its dramatic growth in the case of HNC, which leads
to the no-solution and spinodal-like scenario often spuriously encountered
in the numerical solution of these approximate closures. We will discuss
the occurrence of this type of scenario in the present case, below
in the results sub-sections.

Finally, we have equally examined the Percus-Yevick (PY) closure,
which is a linearised version of HNC

\begin{equation}
g_{i_{\alpha}j_{\beta}}(r)=\exp\left(-\beta v_{i_{\alpha}j_{\beta}}^{(SR)}(r)\right)\left[1++h_{i_{\alpha}j_{\beta}}(r)-c_{i_{\alpha}j_{\beta}}^{(SR)}(r)\right]\label{PY}
\end{equation}
where the superscript (SR) indicates that only the short range interactions
and direct correlations are considered. The long range Coulomb part
is removed from both functions following a standard procedure described
in the literature\cite{74Ng}. In fact, this is also the case in the
HNC and KH closures, because of the exact limiting relation:
\[
\lim_{r\rightarrow\infty}c_{i_{\alpha}j_{\beta}}(r)=-\beta v_{i_{\alpha}j_{\beta}}(r)
\]
which allows to retain only the short range part of the direct correlation
function in the various closure equations. 

As shown in the Results section, both the HNC and the PY equations
cannot be solved for the mixture case, precisely because of the behaviour
of the term $h_{i_{\alpha}j_{\beta}}(r)-c_{i_{\alpha}j_{\beta}}^{(SR)}(r)$,
which induces the growth of medium range correlations, which, in turn,
promotes spurious phase separation-like behaviour. The damping procedure
of the KH closure appears then as a wise method. However, we will
discuss below the appropriateness of this artificial method in view
of the physical phenomena in hydrogen bonding mixtures.

Both the SSOZ equation and the closure equations are solved iteratively
for the unknown site-site functions $h_{i_{\alpha}j_{\beta}}(r)$
and $c_{i_{\alpha}j_{\beta}}(r)$. These functions are discretised
on a grid of 2048 points, with r-spacing of $\Delta r=0.03$\AA. 

We find that the KH closure yields numerical solution for the aqueous
DMSO mixtures for all concentrations at room temperature. Conversely,
neither the PY nor the HNC closures are able to yield numerical solutions
at $T=300$K. While we do not expect that the linearised PY closure
to be able to handle complex structures, the HNC closure deserves
some comments.

The strict HNC closure is unable to yeald numerical solution below
very high temperatures around $T=5000$K for the equimolar mixture.
At such elevated temperature, HNC predicts high small-k raise of the
structure factors, although no obvious long range tail seem to develop
in the site-site correlations. This is a pathology of HNC closure,
as noted early in the historical development of IET\cite{74hncproblem}.
The mathematical origin of such behaviour is unclear at present. It
is generally admitted that this behaviour underlies a spurious spinodal\cite{75hncProblem}.
However, as we have mentioned above, we do not observe that any corresponding
long range tail develops. At such high temperature where we can find
numerical solutions, the site-site functions are nearly structureless,
just like expected for any thermally disordered fluid. In fact, comparing
with the KH closure results at the same temperature reveals that the
first neighbour water-water correlations are overestimated in the
HNC closure, leading to the $k=0$ increase. In other words, it is
the short range behaviour of the correlation which leads to the $k=0$,
and not a spinodal behaviour, contrary to the generally accepted idea
about this closure.

\subsection{Domain order and the scattering function}

The X-ray scattering intensity $I(k)$ can be obtained through the
Debye formula\cite{76DebyeScatt}
\begin{equation}
I(k)=<\sum_{i_{\alpha},j_{\beta}}f_{i_{\alpha}}(k)f_{j_{\beta}}(k)\exp\left(i{\bf k}.({\bf r}_{i_{\alpha}}-{\bf r}_{j_{\beta}})\right)>\label{Idebye}
\end{equation}
where the sum runs over all pairs of scattering atoms $i_{\alpha}$
and $j_{\alpha}$ , which are at respective spatial positions ${\bf r}_{i_{\alpha}}$
and ${\bf r}_{j_{\beta}}$, the functions $f_{i_{\alpha}}(k)$ are
the atomic form factor for atom $i_{\alpha}$ and depend on the type
of radiation which is scattered (we use here Xray), and the symbol
<...> designates an average over all possible positions of these atoms,
which corresponds to a thermal average, or an ensemble average for
calculational purposes. It can be shown\cite{55myDomOrd,76PingsWaser}
that this expression can be rewritten in a form which can be used
in theoretical methods

\begin{equation}
I(k)=I_{\mbox{Ideal}}(k)+I_{\mbox{Intra}}(k)+I_{\mbox{Inter}}(k)\label{I-detail}
\end{equation}
where the 3 contribution represent, respectively the ideal part

\begin{equation}
I_{\mbox{Ideal}}(k)=\sum_{\alpha}x_{\alpha}\sum_{i_{\alpha}}f_{i_{\alpha}}(k)\label{Ideal}
\end{equation}
the intra-molecular part

\begin{equation}
I_{\mbox{Intra}}(k)=\sum_{\alpha\beta}\sqrt{x_{\alpha}x_{\beta}}\sum_{i_{\alpha}j_{\beta}}f_{i_{\alpha}}(k)f_{j_{\beta}}(k)W_{i_{\alpha}j_{\beta}}(k)(1-\delta_{i_{\alpha}j_{\alpha}}\delta_{\alpha\beta})\label{Intra}
\end{equation}
where $W_{i_{\alpha}j_{\beta}}(k)$ is given by Eq.(\ref{W}), and
inter-molecular part

\begin{equation}
I_{\mbox{Inter}}(k)=\sum_{\alpha\beta}\sqrt{x_{\alpha}x_{\beta}}\sum_{i_{\alpha}j_{\beta}}f_{i_{\alpha}}(k)f_{j_{\beta}}(k)H_{i_{\alpha}j_{\beta}}(k)\label{Inter}
\end{equation}
The Kronecker delta terms in Eq.(\ref{Intra}) avoid the self-atomic
contributions inside same species; in which case it is the ideal term
in Eq.(\ref{Ideal}) which prevails. The ideal contribution corresponds
to non-interacting and non-bonded atoms. The ``intra'' part brings
in the fact that atoms are bonded into molecules. Finally the ``inter
`` part accounts for the correlations, hence interaction between
atoms and molecules.

The Xray intensity, as written through Eqs.(\ref{I-detail}-\ref{Inter}),
can be calculated for both the cases of simulations and IET, since
the correlation functions $g_{i_{\alpha}j_{\beta}}(r)$ are available.

\section{Results}

\subsection{Chain-like water clusters}

As shown in Ref.\cite{45ourDMSO1}, the computer simulations of different
combinations of model water and model DMSO indicate that water tends
to form predominantly chain-like clusters. This is illustrated in
Fig.1. This feature is not the one which is usually reported in the
literature, where the accent is more on the water-DMSO dimer formation.
However, these two properties are not necessarily excluding each other.
What is interesting is to know which property induces the other. One
view point is that, water tends to bind preferentially to water, which
then dictates subsequently all other structural features. This property
is reflected in the model simulation through the fact that the partial
charges on the oxygen of water are quite high ($q_{O_{W}}\approx-0.85$
for the oxygen atom, for almost all water models), and higher than
for most solutes (for DMSO, the charge on the oxygen atom is $q_{O_{D}}\approx-0.46$
and on the sulfur atom $q_{S_{D}}\approx0.14$, for acetone $q_{O_{A}}\approx-0.56,$
and for mono-ols $q_{O}\approx-0.7$). A logical consequence of electrostatic
is that charges with highest valence will tend to dominate the interactions,
leading weaker charges to adapt. Following this line of reasoning,
water is more likely to bind with itself, and preserve pockets of
tetrahedrally bound water molecules. This is what happens with mono-ols\cite{77ourtba}
and acetone\cite{78ourace}. We would then expect something similar
with DMSO as well, and at even greater scale, since the valence is
smaller. This is not what sis observed in Fig.1: water seems to mix
rather well with DMSO, precisely by forming chain-like clusters. Clearly,
the competition between charges alters the simplistic picture presented
above. High valences on the solute equally favours self solute binding,
thereby enforcing the micro-segregation observed in simulation of
aqueous-acetone\cite{78ourace} or aqueous-alcohol\cite{55myDomOrd,77ourtba}.
Intuitively, weaker charges on the solute should increase water segregation.
This is what we observed in aqueous mixtures of ``weak-water'' models,
where the valence of the latter weak-water was decreased to various
values. Therefore, the reason for the enhanced miscibility of water
in these mixtures is to be searched elsewhere.

\subsection{Structure functions }

We look at the site-site correlation functions and associated structure
factors of pure DMSO and aqueous DMSO, as obtained from computer simulations
and IET. 

\subsubsection{Pure water}

The case of pure water has been reported in several previous works
by many authors\cite{57LueBlanck1,61a,61b}, including our own work\cite{73ourism}.
The comparison with scattering experiments shows that classical force
field such as SPC/E are quite adequate. However, structure prediction
from IET cannot capture the strong tetrahedral ordering of water,
and predict a structure closer to that of a simple Lennard-Jones liquid.
Interestingly, and as stated above, both the molecular and the site-site
produce very similar correlations, which indicates that the principal
issue is not in the differences between the two methods, but the fact
\emph{they both miss many body correlations}. In the present work,
the water-water correlations for mixtures are corrected by the pure
water bridge terms, which we have computed in Ref.\cite{73ourism}.
We show that this contribution is capital to produce the chain-like
water clusters within the HNC theory.

\subsubsection{Neat DMSO}

Fig.2 shows selected atom-atom correlation functions and corresponding
structure factors for pure DMSO, as obtained from simulations (curve
in green), HNC (blue), KH(red dashes) and PY(magenta). It is seen
that the first neighbor correlations are generally underestimated
by the theories, but not dramatically so. The PY theory is not able
to properly describe the features of the structure factors. But the
principal problem with the theories is seen in small-k behaviour of
the structure factors, which all show a pronounced raise, suggesting
liquid-vapour pre-transitional prediction. This is most serious for
the HNC closure. A close look at the long range behaviour of the correlation
functions shows a general tendency to be above the asymptote 1.

\subsubsection{Aqueous-mixtures}

Selected correlations between the oxygen atoms of water and DMSO molecules,
as well as associated structure factors, are displayed for typical
DMSO mole fractions x=0.1, 0.5 and 0.8, in Fig.3, Fig.4 and Fig.5,
respectively. The color conventions are similar to Fig.2, simulation
results in green, HNC in blue and KH in red dashes. What is immediately
apparent is how poor IET results are. Despite the fact that water-water
correlations contain the bridge function which reproduces pure water
data accurately, it is not sufficient to enforce a better agreement
than the KH closure. In fact, this closure seems to reproduce the
trends of the correlations for the 10\% DMSO case in Fig.4a, while
the enhanced HNC is very poor. The situation is somewhat improved
for the 50\% and 80\% DMSO cases. Perhaps the most important feature
comes from the structure factors in the small-k region, where the
enhanced HNC version tends to predict the cluster pre-peak visible
on the simulation data at all concentrations. The KH approximation,
despite being somewhat more accurate, predicts a small-k raise at
k=0 for the water-water correlations. In fact, the strict HNC closure
cannot be solved for this mixture at room temperature, precisely because
of spurious k=0 raise in the correlations. The linearisation present
in the KH closure prevents this raise to be happen in the same dramatic
fashion that for the strict HNC closure, but it is at the heart of
these approximations limited to pair correlations. It is only through
the introduction of higher order correlation that this problem can
be cured. This fact indicates how crucial such correlations are to
represent the complexity of associating mixtures. 

As indicated in our previous study of aqueous-DMSO mixtures, the most
important feature of the correlations is the presence of a pre-peak
in the water-water structure factors, in the k-vector range $k\approx0.7-1$\AA$^{-1}$.
This is due to chain-like water aggregates, as indicated by the second
and third neighbour depression in the water-water correlations. These
positive pre-peak contributions are accompanied with negative pre-peak
contributions in the water-DMSO cross correlations. This was clearly
illustrated in Fig.7 of Ref.\cite{45ourDMSO1}. The presence of positive
and negative pre-peaks is not specific to the present system, and
are equally found in room temperature ionic liquids\cite{80ourIL}.
They correspond in fact to an underlying charge ordering between positively
and negatively charged atomic groups. Despite the strong local inhomogeneity
induced by such charge order, which is responsible for the micro-heterogeneity
as well as the linear water aggregates found here, the global homogeneity
is preserved. A signature of this global homogeneity is the perfect
compensation of these positive and negative pre-peak in the scattered
intensity, as we show in sub-section 4 below.

\subsection{Charge order}

An important property of the structure of neat DMSO is revealed in
Fig.6. It shows that the long range correlations between the oxygen
and sulfur atoms with methyl atoms are out-of-phase with all the types
of atom-atom the correlations. This property is typical of charge
ordering\cite{80pccpCHORD}, as recently advertised in room temperature
ionic-liquids\cite{81castnerIL,82VothIL}. Charge order essentially
describes the alternated disposition of positive and negatively charged
atoms, most appropriately in ionic melts. But this order can equally
occur between charged atomic groups in complex room temperature ionic
liquids\cite{81castnerIL,82VothIL}. The hydrogen bonding in associative
liquids is such a nice example of charge ordering, alternating oxygen
and hydrogen atoms, which in this case enforce the tetrahedral ordering
in water, or the chain-like ordering in mono-ols. Therefore, what
we observe in DMSO is an alternance of oxygen and methyl. We conjecture
here that it is this ordering which imposes the ordering of the water
molecules, which then are not obliged to cluster into globular clusters,
but rather meander between the DMSO molecules through linearly bound
water chains. Fig.3 equally shows that none of the theories are able
to describe the strength of this charge ordering within the same spatial
range, and considerably underestimate it. This could be due to high
order diagrams missing. We did not try incoporating such diagrams
by the procedure which we used in other cases, for the reason explained
in the next sub-section.

\subsection{Kirkwood-Buff integrals}

Fig.7 shows the KBI from experiments (filled dots) compared with simulations
(open triangles) and IEt results, HNC (full lines) and KH( dashed
lines). We note that generally, the agreement between theoretical
and experimental results are rather good. This is even more surprising
when the IET results for structure are quite disappointing. We note
that the KH results are stunning accurate compared to the experiments,
except for $G_{DD}$for small DMSO content. The enhanced HNC results
are worse for $G_{WW}$ at high DMSO content. In fact, they seem to
follow the results of the simulations rather well, which is intriguing.
Indeed, the force field model we have used are exactly the same for
both theoretical methods. Despite these shortcomings, this thermodynamic
property related to structure is rather well reproduced. This is in
sharp contrast with the KBI results for other aqueous mixtures, such
as water- alcohol, for example, for which the KBI from IET are very
poor. Comparing the nature of the aggregation in both systems, since
aqueous alcohol mixtures tend to present quite large water-water KBI
in relation with strong micro-heterogeneity and globular water aggregates,
while aqueous DMSO has rather small micro-heterogeneity and linear
water aggregates, it would seem that IET are better suited for weakly
aggregating mixtures. This is inline with our previous work with aqueous
weak-water mixture models\cite{91weakwater3}.

\subsection{Scattering function}

Fig.8a-b show the Xray scattering intensity from all 3 theoretical
methods compared with experimental data for pure water and pure DMSO.
For water, the full intensity is represented. For DMSO, another common
representation is often used by experimentalists and simulators is:

\begin{equation}
\Delta I(k)=k.\left(I(k)-I_{\mbox{ideal}}(k)\right)\label{DI(k)}
\end{equation}
It can be seen that, despite minor discrepancies in the correlation
functions, the IET manage to reproduce the total scattered intensity
rather well.

Fig.9a-c show the scattering intensity for mixtures, compares with
the Xray data from Nishikawa\cite{50KogaXray}. We note that the simulations
are in good agreement with the experimental data, and that the IET
perform quite well, specially when compared with the rather poor structure
factors shown in Figs.4. This is certainly due to favourable compensations
in the sum of Eq.(\ref{Inter}).

Perhaps the most important feature is the disappearance of the water
pre-peak from the total intensity. This is due to a perfect compensation
from the water-water and water-DMSO cross contributions in Eq.(\ref{Inter}).
The positive pre-peak of the first are exactly compensated by the
negative contributions from the second. This can be seen in the partial
contributions shown in the insets.

The insets of the figures show the total scattered intensity $I(k)$,
together with other contributions in Eq.(\ref{I-detail}), namely
$I_{\mbox{ideal}}(k)$ in green line and $\Delta I_{\mbox{Intra}}(k)=I_{\mbox{Intra}}(k)-I_{\mbox{Ideal}}(k)$
in jade. The water-water and water-DMSO cross contributions are equally
shown in cyan and magenta, respectively. These 2 curves serve to illustrate
how the water aggregate pre-peak gets compensated into the total intensity
I(k) (in blue). It is equally important to realize that the intra-molecular
contribution is quite important, and needs to be added to the intermolecular
contribution which is solely made of the site-site functions.

\section{Discussion and Conclusion}

The present work illustrates how approximate IET, which are obviously
unable to account for the local ordering due to association in aqueous
mixtures, can however turn out to be relatively acceptable when computing
averaged quantities accessible to experiments, such as the KBI or
the radiation scattering intensity. The first concerns a spatial average,
and is related to various thermodynamic properties\cite{95KB} , while
the second is an average through summing over all site-site contributions.
IET are relatively accurate in many situations, such as simple liquids
and their mixtures\cite{97a}, weakly polar systems\cite{97b}, ionic
melts\cite{97c}. The present results indicate that their accuracy
comes essentially when 2-body interactions and correlations are predominant.
If clustering is present, then these theories become useless. A supplementary
hint comes from the fact that IET are very unreliable to describe
critical fluctuations, which means in the vicinity of phase transitions\cite{75hncProblem}.
The HNC theory is not even able to predict a spinodal behaviour\cite{75hncProblem}.
Since aggregation is a k-dependent concentration fluctuation, the
latter which simply the k=0 part, we do not expect these theories
to be able to describe \emph{directional clustering} properly. This
is the principal reason of the problems reported here. Nevertheless,
the fact that they can be used for weakly aggregating systems, such
as aqueous-DMSO, indicate the importance of many body correlations
in various other aqueous mixtures. The multiple recipe to improve
IET for spherical interaction are of no use when directional clustering
is present. The present work, equally indicates that accounting for
pure water association is not enough to permit to describe what happens
in the case of mixtures. The no-spinodal pathology of the HNC closure
indicates that recepy such as the KH closure do not bring any deep
insight into the relation between directional molecular association
and many body correlations. Insights into this matter are still missing.

It is equally interesting to see that radiation scattering, ore specifically
Xray scattering, does not account for the local structure of the aqueous-DMSO
mixtures, namely the linear chain aggregates. While this is quite
apparent from the water-water correlations in many model combinations,
it is not immediately deducible from spectroscopy data, which tends
to enhance water-DMSO pairing, and even less from scattering data,
where the pre-peak is totally absent. In a way, this result indicates
that computer simulations provide a better insight into the local
microscopic structure than experimental radiation scattering can provide.
Perhaps neutron scattering experiments can provide additional insight
into this matter, through various deuteration processes. This point
remains open for subsequent investigations.

\newpage

\section*{Figure captions}
\begin{lyxlist}{00.00.0000}
\item [{Fig.1}] Comparison between various water-water RDF reported in
the literature for aqueous-DMSO mixtures. Blue line is pure SPC/E
water, red line is x=0.35 from Ref.\cite{32chandlerLuzar2}, green
line for x=0.33 from Ref.\cite{36Laaksonen}, jade line is for x=0.3
from Ref.\cite{45ourDMSO1}, orange line from OPLS DMSO, and dotted
gray line for aqueous ethanol x=0.3\cite{99marijana}.
\item [{Fig.}] Snapshot of the 50\% aqueous DMSO mixture (2048 particles)
evidencing linear water clusters. Water molecules are shown as red
ball for oxygen atom and white for hydrogen atoms. DMSO is shown in
cyan semi-transparent.
\item [{Fig.3}] Selected atom-atom correlation functions (a) and corresponding
structure factors (b) for the neat DMSO mixture, comparing simulation
data (green) to HNC(blue), KH(dashed red) and PY(thin magenta) data.
The various atom pairs are indicated in the panels.
\item [{Fig.4}] Oxygen atom correlation functions (a) and associated structure
factors (b) for the 10\% aqueous DMSO mixture. The color conventions
are as in Fig.2.
\item [{Fig.5}] Oxygen atom correlation functions (a) and associated structure
factors (b) for the 50\% aqueous DMSO mixture. The color conventions
are as in Fig.2.
\item [{Fig.6}] Oxygen atom correlation functions (a) and associated structure
factors (b) for the 80\% aqueous DMSO mixture. The color conventions
are as in Fig.2.
\item [{Fig.7}] DMSO atom-atom correlation functions demonstrating the
long range charge order behaviour (see text). $g_{OO}(r)$ in blue,
$g_{SS}(r)$ in purple, $g_{SM}(r)$ in black, $g_{MM}(r)$ in green,
$g_{OS}(r)$ in red and $g_{OM}(r)$ in orange. The main panel is
for neat DMSO, the upper two insets for 80\% and 50\% aqueous DMSO
mixtures, and the lower inset for the HNC results.
\item [{Fig.8}] Kirkwood-Buff integrals of the aqueous-DMSO mixtures. $G_{WW}$
in blue, $G_{DD}$ in green and $G_{WD}$ in magenta. Filled dots
are experimental results from Ref.\cite{65Matteoli}, open triangles
simulation results from Ref.\cite{45ourDMSO1}, blue curve for HNC
and dashed curves for KH.
\item [{Fig.9}] Xray scattering intensities $I(k)$ for neat water (a)
and neat DMSO (b), together with the $\Delta I(k)$ representations
from Eq.(\ref{DI(k)}) Experimental results in red, simulation in
blue, IET results in dotted cyan. The ideal contribution from Eq.(\ref{Ideal})
is in green and the intra-molecular from Eq.(\ref{Intra}) contribution
in jade. For water, the experimental data for $I(k)$ is from Ref.\cite{98XrayWater},
and $\Delta I(k)$ for both water and DMSO are from from Ref.\cite{50KogaXray}.
\item [{Fig.10}] Xray scattering intensities $\Delta I(k)$ (main panel)
and $I(k)$ (inset) for aqueous mixtures, with 17\% (a), 50\% (b)
and 80\% (c) DMSO contents. Experimental results (from Ref.\cite{50KogaXray})
in red, simulation in blue, HNC results in cyan, and KH in dashed
jade. The ideal and intra-molecular contributions are shown in the
insets in dashed green and jade, respectively. The purple curve represents
the water-water contribution to I(k), while the magenta curve is for
the cross water-DMSO contribution (see text)
\end{lyxlist}
\newpage

.

\includegraphics[scale=0.6]{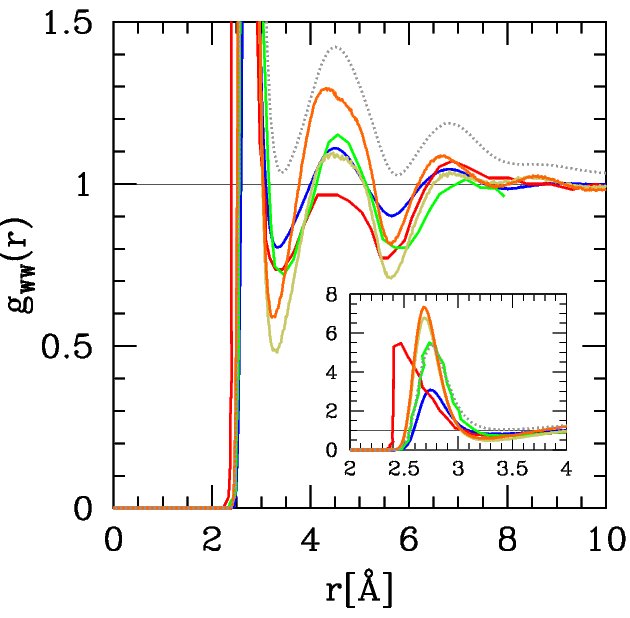}

.

.Fig.1 - Comparison between various water-water RDF reported in the
literature for aqueous-DMSO mixtures. Blue line is pure SPC/E water,
red line is x=0.35 from Ref.\cite{32chandlerLuzar2}, green line for
x=0.33 from Ref.\cite{36Laaksonen}, jade line is for x=0.3 from Ref.\cite{45ourDMSO1},
orange line from OPLS model\cite{67oplsDMSO}, and dotted gray line
for aqueous ethanol x=0.3\cite{99marijana}.

\newpage

.

\includegraphics[scale=0.5]{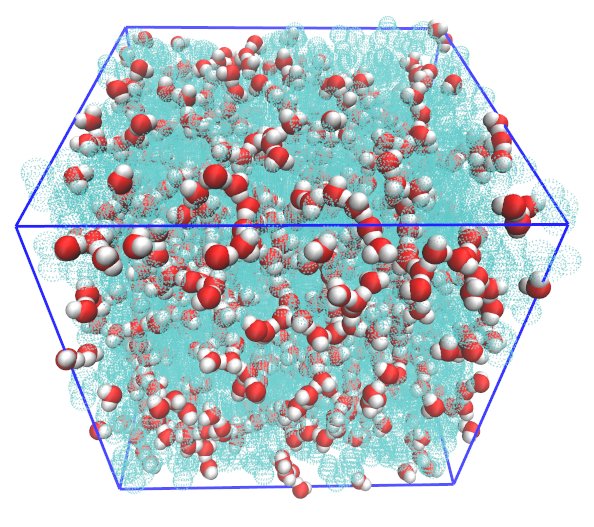}

.

.Fig.2 - Snapshot of the 50\% aqueous DMSO mixture (2048 particles)
evidencing linear water clusters. Water molecules are shown as red
ball for oxygen atom and white for hydrogen atoms. DMSO is shown in
cyan semi-transparent.

.

\newpage

.

\includegraphics[scale=0.4]{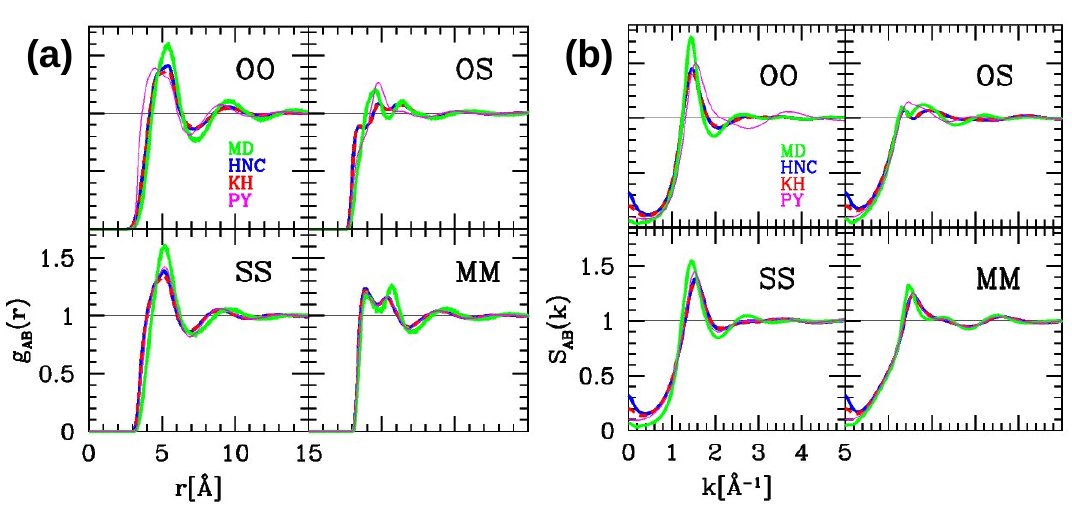}

.

.Fig.3 - Selected atom-atom correlation functions (a) and corresponding
structure factors (b) for the neat DMSO mixture, comparing simulation
data (green) to HNC(blue), KH(dashed red) and PY(thin magenta) data.
The various atom pairs are indicated in the panels.

.

\newpage

.

\includegraphics[scale=0.4]{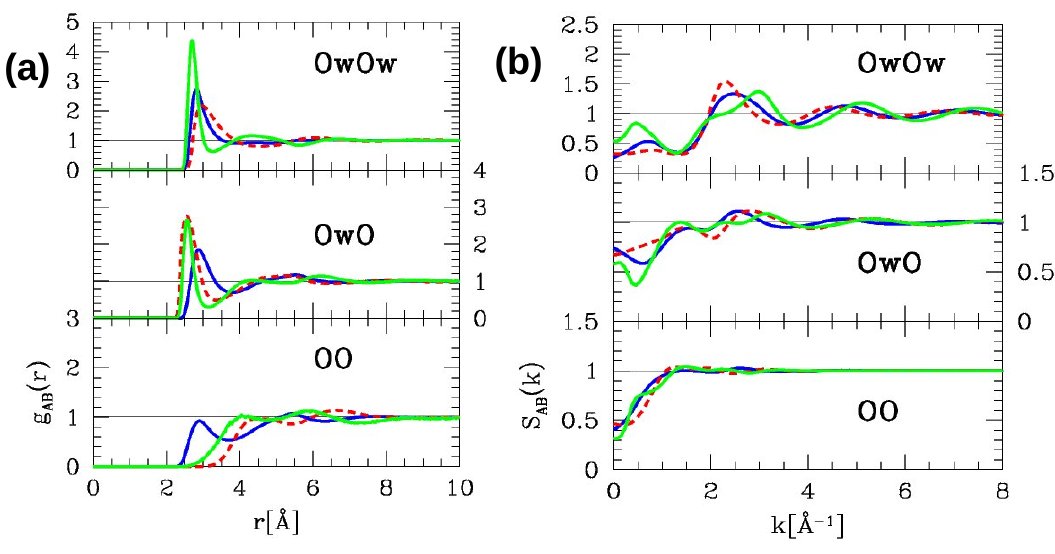}

.

.Fig.4 - Oxygen atom correlation functions (a) and associated structure
factors (b) for the 10\% aqueous DMSO mixture. The color conventions
are as in Fig.2.

.

.\newpage

.

\includegraphics[scale=0.4]{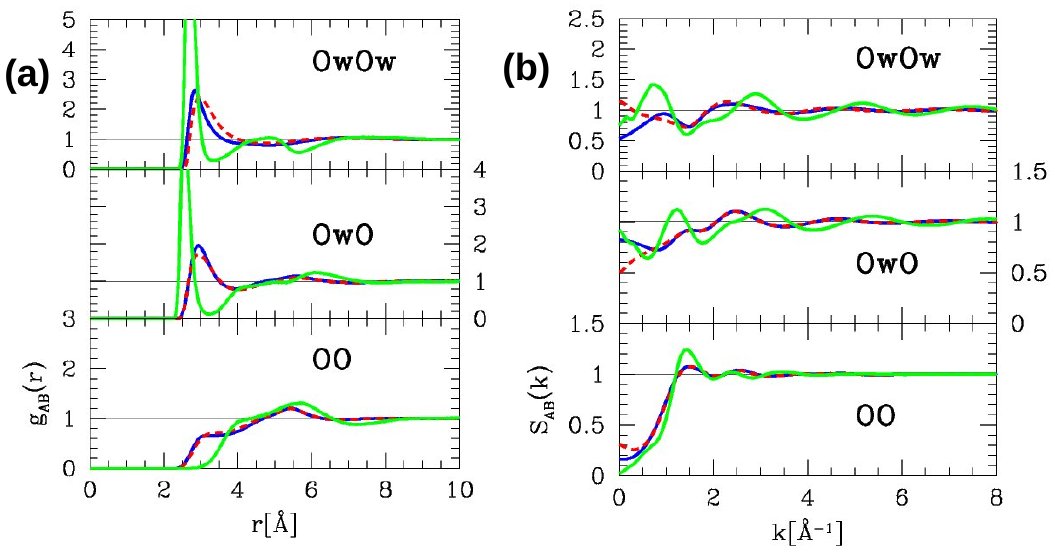}

.

.Fig.5 - Oxygen atom correlation functions (a) and associated structure
factors (b) for the 50\% aqueous DMSO mixture. The color conventions
are as in Fig.2.

.

.\newpage

.

\includegraphics[scale=0.4]{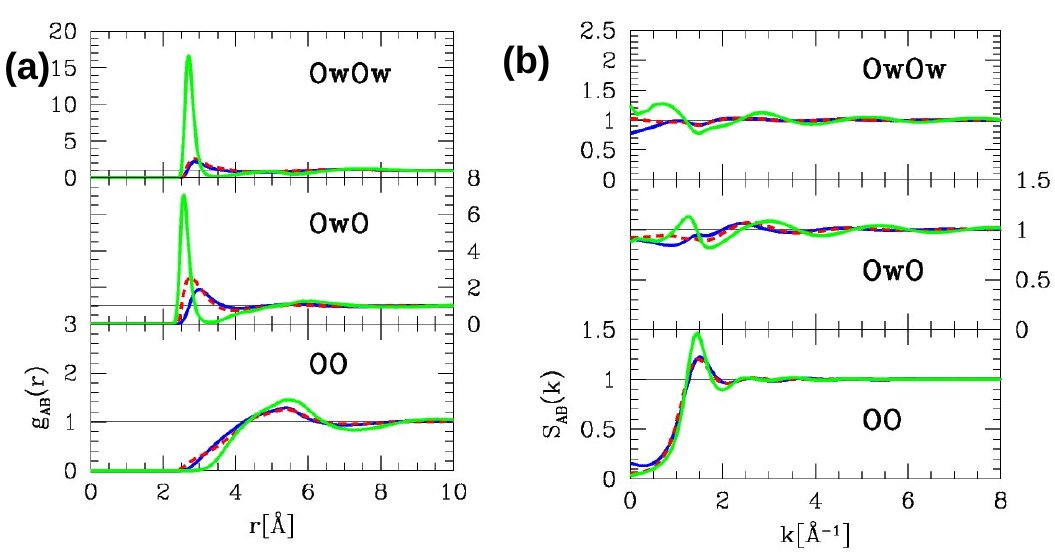}

.

.Fig.6 - Oxygen atom correlation functions (a) and associated structure
factors (b) for the 80\% aqueous DMSO mixture. The color conventions
are as in Fig.2.

.

.\newpage

.

\includegraphics[scale=0.5]{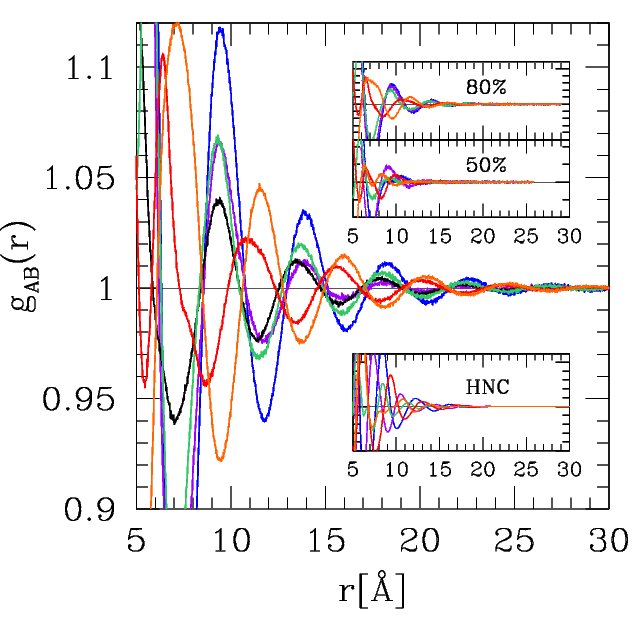}

.

.Fig.7 - DMSO atom-atom correlation functions demonstrating the long
range charge order behaviour (see text). $g_{OO}(r)$ in blue, $g_{SS}(r)$
in purple, $g_{SM}(r)$ in black, $g_{MM}(r)$ in green, $g_{OS}(r)$
in red and $g_{OM}(r)$ in orange. The main panel is for neat DMSO,
the upper two insets for 80\% and 50\% aqueous DMSO mixtures, and
the lower inset for the HNC results.

.\newpage

.

\includegraphics[scale=0.5]{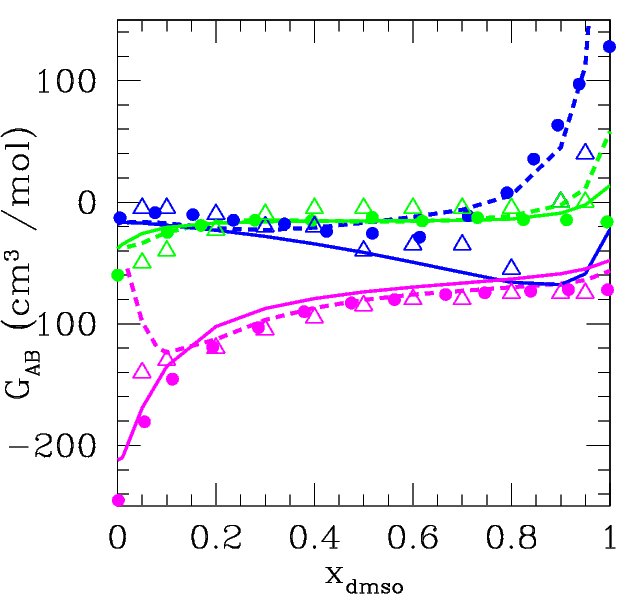}

.

.Fig.8 - Kirkwood-Buff integrals of the aqueous-DMSO mixtures. $G_{WW}$
in blue, $G_{DD}$ in green and $G_{WD}$ in magenta. Filled dots
are experimental results from Ref.\cite{65Matteoli}, open triangles
simulation results from Ref.\cite{45ourDMSO1}, blue curve for HNC
and dashed curves for KH.

.\newpage

.

\includegraphics[scale=0.4]{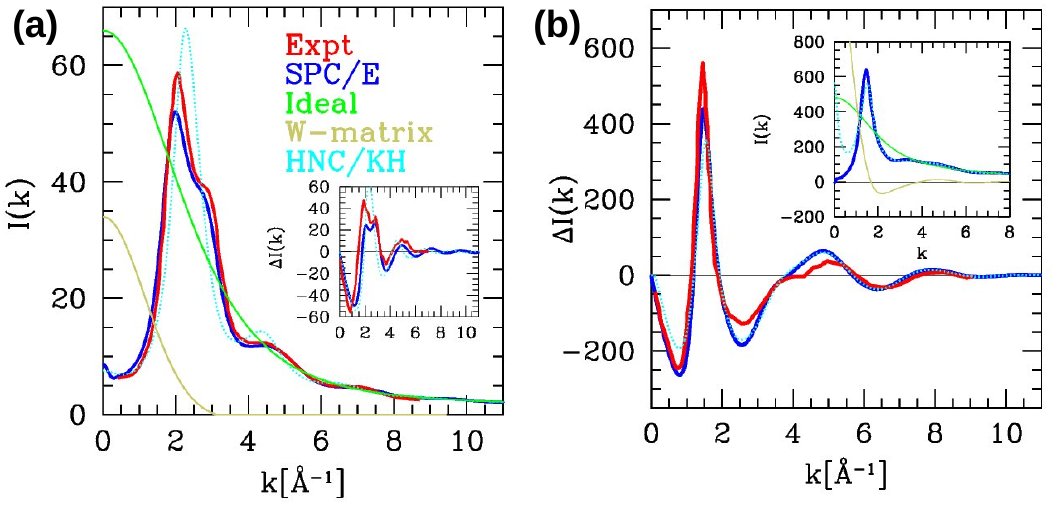}

.

.Fig.9 - Xray scattering intensities $I(k)$ for neat water (a) and
neat DMSO (b), together with the $\Delta I(k)$ representations from
Eq.(\ref{DI(k)}) Experimental results in red, simulation in blue,
IET results in dotted cyan. The ideal contribution from Eq.(\ref{Ideal})
is in green and the intra-molecular from Eq.(\ref{Intra}) contribution
in jade. For water, the experimental data for $I(k)$ is from Ref.\cite{98XrayWater},
and $\Delta I(k)$ for both water and DMSO are from from Ref.\cite{50KogaXray}..

.

.\newpage

.

\includegraphics[scale=0.4]{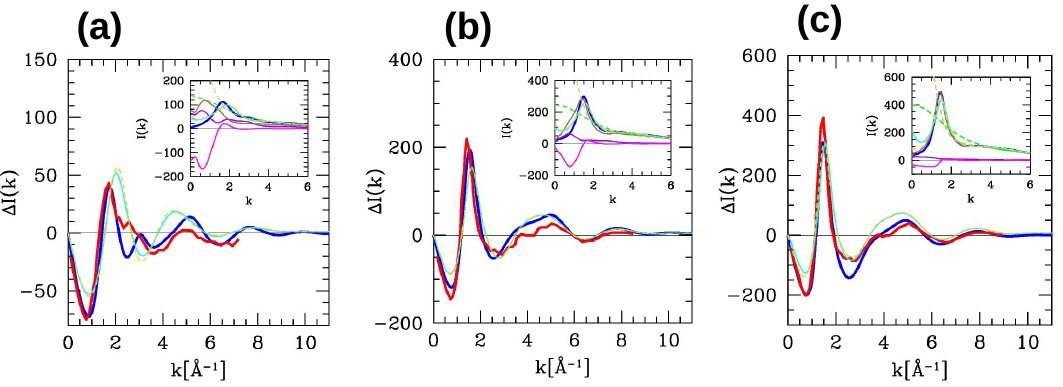}

.

.Fig.10 - Xray scattering intensities $\Delta I(k)$ (main panel)
and $I(k)$ (inset) for aqueous mixtures, with 17\% (a), 50\% (b)
and 80\% (c) DMSO contents. Experimental results (from Ref.\cite{50KogaXray})
in red, simulation in blue, HNC results in cyan, and KH in dashed
jade. The ideal and intra-molecular contributions are shown in the
insets in dashed green and jade, respectively. The purple curve represents
the water-water contribution to I(k), while the magenta curve is for
the cross water-DMSO contribution (see text)) 

.
\end{document}